\documentclass[a4paper,11pt]{article}
\pdfoutput=1 

\usepackage{jcappub} 

\usepackage{xcolor}
\usepackage{hyperref}
\usepackage{amsmath}
\usepackage{multirow}
\usepackage{graphicx}
\usepackage{dcolumn}
\usepackage{bm}
\usepackage{amssymb}
\usepackage{latexsym}
\usepackage{booktabs}
\usepackage{amsmath}
\usepackage{enumerate}
\usepackage{url}
\usepackage{subfigure}
\usepackage{afterpage}
\usepackage{placeins}
\usepackage{float}

\begin{document}

\title{Nanohertz gravitational waves from a quasar-based supermassive black hole binary population model as dark sirens}

\author[a]{Si-Ren Xiao,}
\author[a]{Yue Shao,}
\author[b]{Ling-Feng Wang,}
\author[a]{Ji-Yu Song,}
\author[c,a]{Lu Feng,}
\author[a]{Jing-Fei Zhang}
\author[a,d,e,\ast]{and Xin Zhang}\note[$\ast$]{Corresponding author.}

\affiliation[a]{Liaoning Key Laboratory of Cosmology and Astrophysics, College of Sciences, Northeastern University, Shenyang 110819, China}
\affiliation[b]{School of Physics and Optoelectronic Engineering, Hainan University, Haikou 570228, China}
\affiliation[c]{College of Physical Science and Technology, Shenyang Normal University, Shenyang 110034, China}
\affiliation[d]{MOE Key Laboratory of Data Analytics and Optimization for Smart Industry, Northeastern University, Shenyang 110819, China}
\affiliation[e]{National Frontiers Science Center for Industrial Intelligence and Systems Optimization, Northeastern University, Shenyang 110819, China}
\emailAdd{xiaosiren@stumail.neu.edu.cn, shaoyue@stumail.neu.edu.cn, wanglf@hainanu.edu.cn, songjiyu@stumail.neu.edu.cn, fengluu@foxmail.com, jfzhang@mail.neu.edu.cn, zhangxin@mail.neu.edu.cn}

\abstract{
Recently, several pulsar timing array (PTA) projects have detected evidence of the existence of a stochastic gravitational wave background (SGWB) in the nanohertz frequency band, providing confidence in detecting individual supermassive black hole binaries (SMBHBs) in the future. Nanohertz GWs emitted by inspiraling SMBHBs encode the luminosity distances of SMBHBs. They can serve as dark sirens to explore the cosmic expansion history via a statistical method to obtain the redshift information of GW sources' host galaxies using galaxy catalogs. The theoretical analysis of the dark siren method relies on the modeling of the population of SMBHBs. Using a population model consistent with the latest SGWB observations is essential, as the SGWB provides significant information about the distribution of SMBHBs. In this work, we employ a quasar-based model, which can self-consistently account for the SGWB amplitude, to estimate the population of SMBHBs. We constrain the Hubble constant using the mock GW data from different detection cases of PTAs in the future. 
Our results show that a PTA consisting of 100 pulsars with a white noise level of 20 ns could measure the Hubble constant with a precision close to 1\% over a 10-year observation period, and a PTA with 200 pulsars may achieve this goal over a 5-year observation period. The results indicate that modeling the SMBHB population significantly influences the analysis of dark sirens, and SMBHB dark sirens have the potential to be developed as a valuable cosmological probe.
}

\maketitle
\section{Introduction} 
\label{sec:intro}

The study of cosmology has entered the era of precision cosmology \cite{WMAP:2003elm,WMAP:2003ivt}. 
Five of the six basic parameters of the $\Lambda$CDM model have been precisely constrained by the \emph{Planck} cosmic microwave background (CMB) data with constraint precisions better than 1\% \cite{Planck:2018vyg}.
However, as the precision of cosmological parameter measurements improves, some puzzling tensions have come to light. The Hubble constant ($H_0$) is a crucial cosmological parameter describing the universe's current expansion rate. Its values inferred from the \emph{Planck} CMB observation assuming the $\Lambda$CDM model~\cite{Planck:2018vyg} and the model-independent local-universe distance ladder measurement \cite{Riess:2021jrx} are in more than $5\sigma$ tension, which has become a critical crisis in cosmology \cite{Verde:2019ivm,Riess:2019qba} and has been widely discussed (see, e.g.~\cite{Li:2013dha,Zhang:2014ifa,Zhang:2014nta,Zhang:2014dxk,Zhao:2017urm,Guo:2017qjt,Guo:2018ans,Yang:2018euj,Verde:2019ivm,Riess:2019qba,Guo:2019dui,cai:2020,DiValentino:2019jae,DiValentino:2019ffd,Liu:2019awo,Zhang:2019cww,Ding:2019mmw,Cai:2021wgv,DiValentino:2021izs,Abdalla:2022yfr,Li:2020tds,Wang:2021kxc,Vagnozzi:2021tjv,Vagnozzi:2021gjh,Vagnozzi:2019ezj,Feng:2019jqa,Lin:2020jcb,Hryczuk:2020jhi,Gao:2021xnk,Zhao:2022yiv,Gao:2022ahg}). 
While the Hubble tension indicates the potential for new physics beyond the $\Lambda$CDM model, there remains a lack of consensus on a valid extended cosmological model to address this tension. Additionally, it is imperative to explore new cosmological probes capable of independently measuring $H_0$.

Compact binary coalescence (CBC) events, a category of gravitational wave (GW) sources, have been discovered by the LIGO-Virgo-KAGRA collaboration in recent years, with approximately a hundred events detected.
The luminosity distances of CBC events can be determined through the analysis of GW waveforms, and these events are commonly referred to as ``standard sirens'' \cite{Schutz:1986gp, Holz:2005df}. 
Standard sirens can provide absolute distance measurements, avoiding poorly understood calibration processes inherent in the distance ladder method. If the redshifts of GW sources are also determined, the distance-redshift relation could be established to study the expansion history of the universe (see, e.g.~\cite{Dalal:2006qt, Cutler:2009qv,Zhao:2010sz, Cai:2016sby,Cai:2017plb, Cai:2017aea,Zhao:2018gwk,Wang:2018lun,Zhang:2018byx, Zhang:2019ple,Li:2019ajo, Cai:2018rzd,Chang:2019xcb, Du:2018tia, He:2019dhl, Yang:2019vni, Yang:2019bpr, Zhang:2019ylr, Bachega:2019fki,Zhang:2019loq,Chen:2020dyt,Jin:2020hmc, Chen:2020zoq,Mitra:2020vzq,Zhao:2020ole,Jin:2021pcv,Jin:2022tdf, Jin:2022qnj,Jin:2023zhi, Jin:2023tou,Jin:2023sfc, Qi:2021iic,Ye:2021klk, Cao:2021zpf,deSouza:2021xtg, Wang:2021srv,Colgain:2022xcz, Dhani:2022ulg,Wang:2022rvf, Zhu:2021bpp,Califano:2022syd, Hou:2022rvk,Wu:2022dgy, Han:2023exn,Zhang:2023gye, Li:2023gtu,Yu:2023ico}). 
The binary neutron star coalescence event, GW170817, gives $H_0=70^{+12}_{-8}~\rm~km~s^{-1}~Mpc^{-1}$ with a precision of 14\% \cite{LIGOScientific:2017adf}. Its redshift was determined by directly identifying the electromagnetic (EM) signals produced when the binary merged \cite{LIGOScientific:2017vwq, LIGOScientific:2017ync}. Such standard sirens are known as bright sirens, with only one case detected to date. 

For the standard sirens without EM counterparts (usually referred to as dark sirens), we can cross-match them with galaxy catalogs and statistically infer their redshifts (see, e.g.~\cite{DelPozzo:2011vcw,Chen:2017rfc,Nair:2018ign,LIGOScientific:2018gmd,Gray:2019ksv,DES:2019ccw,LIGOScientific:2019zcs,Yu:2020vyy,DES:2020nay,Palmese:2021mjm,LIGOScientific:2021aug,Leandro:2021qlc,Finke:2021aom,Gair:2022zsa,Yang:2022iwn,Muttoni:2023prw,Song:2022siz}). Recently, using 46 dark sirens with signal-to-noise ratios (SNRs) over 11 from GWTC-3 and the GLADE+ $K$-band galaxy catalog \cite{Dalya:2018cnd, Dalya:2021ewn}, researchers achieved a $\sim$19\% constraint on $H_0$. Combined with the bright siren GW170817, the constraint precision of $H_0$ improves to $\sim$10\% \cite{LIGOScientific:2021aug}. 

The frequencies of GWs generated by the GW sources with varying masses are hugely different, which requires GW detectors sensitive to diverse frequency bands. The current LIGO-Virgo-KAGRA detector network \cite{Somiya:2011np, LIGOScientific:2014pky, VIRGO:2014yos} can detect only stellar-mass CBCs at around hundreds of hertz. In the future, precise measurements of GWs in various bands will become possible. In the frequency band from a few hertz to several thousand hertz, the third-generation ground-based GW detectors, e.g., the Cosmic Explorer \cite{LIGOScientific:2016wof} and the Einstein Telescope \cite{Punturo:2010zz}, will significantly enhance the detection capabilities for stellar-mass CBCs. In the decihertz band, the Decihertz Interferometer Gravitational-Wave Observatory \cite{Seto:2001qf, Kawamura:2018esd,Dong:2024bvw} and lunar-based detectors \cite{LGWA:2020mma, Jani_2021} could detect intermediate-mass black hole binaries and binary white dwarfs. In the millihertz band, GWs from massive black hole binaries and extreme mass ratio inspirals could be detected by space-borne GW detectors such as the Laser Interferometer Space Antenna \cite{LISA:2017pwj, Robson:2018ifk, LISACosmologyWorkingGroup:2022jok}, Taiji \cite{Wu:2018clg,Ruan:2018tsw,Hu:2017mde,Zhao:2019gyk}, and TianQin~\cite{TianQin:2015yph,Wang:2019ryf,wang:2019tto,Luo:2020bls,Milyukov:2020kyg,TianQin:2020hid}. In the nanohertz (nHz) band, the pulsar timing arrays (PTAs) could detect continuous GWs from the inspiraling supermassive black hole binaries (SMBHBs) resulting from galaxy mergers.

The pulsars constructing PTAs are millisecond pulsars (MSPs) and have highly stable rotation periods. Their radio signals can be periodically detected by radio telescopes on the Earth with extremely high precision. GWs passing between the MSPs and the Earth will alter the proper distances between them. This leads to a discrepancy between the actual and expected arriving times of radio signals, which is known as the timing residuals. By analyzing the time residuals of PTAs, we can extract information about nHz GWs.
Recently, North American Nanohertz Observatory for Gravitational Waves (NANOGrav) \cite{NANOGrav:2023gor, NANOGrav:2023hde}, European PTA (EPTA) + Indian PTA (InPTA) \cite{EPTA:2023fyk, EPTA:2023sfo}, Parkes PTA (PPTA) \cite{Zic:2023gta, Reardon:2023gzh}, and Chinese PTA (CPTA) \cite{Xu:2023wog} have reported substantial evidence for the existence of a stochastic GW background (SGWB) \cite{Hellings:1983fr}.
Such evidence boosts confidence in making more advances in detecting nHz GWs. For instance, detecting individual SMBHBs has become one of the next important missions of PTA observations~\cite{Sesana:2008xk,Corbin:2010kt,Sesana:2010ac,Babak:2011mr,Lee:2011et,Ellis:2012zv,Ellis:2013hna,Zhu:2015tua,Wang:2016tfy,Feng:2020nyw}.
 Although extracting an individual GW signal from the superimposed GW background is challenging, this goal may be achievable with the addition of more advanced radio telescopes, especially in the era of the Square Kilometre Array (SKA).

Recently, Yan et al. \cite{Yan:2020ewq} and Wang et al. \cite{Wang:2022oou} indicated that utilizing SMBHBs as standard sirens could help to improve the measurement precisions of cosmological parameters. 
Yan et al. \cite{Yan:2020ewq} pointed out that the EM counterparts of SMBHBs can be observed during its inspiraling stage and forecast the constraint on the equation of state of dark energy with 154 SMBHB bright sirens observed by SKA-era PTAs.
Wang et al. \cite{Wang:2022oou} investigated the dominant factors affecting the detection capability of PTAs for SMBHBs and found that the root mean square (RMS) of white noise, $\sigma_t$, is more dominant than the number of MSPs in affecting the detections of SMBHBs. They also proposed that detecting nHz GWs from individual SMBHBs has significant implications for exploring the properties of dark energy and precisely determining the Hubble constant.

In the aforementioned papers, the SMBHB standard siren mock data are simulated based on the galactic major merger model proposed by Mingarelli et al. \cite{Mingarelli:2017fbe} (hereafter the M17 model).
Recently, Casey-Clyde et al. \cite{Casey-Clyde:2021xro} indicated that most major merger models predicted characteristic amplitudes of SGWB lower than that derived from the NANOGrav 12.5-yr data. In this context, they proposed a quasar-based SMBHB population model (hereafter the C21 model), considering that galaxy mergers could trigger quasars, and SMBHBs arising from galaxy mergers might be linked to quasar populations. This model can self-consistently predict the local number density of SMBHBs given the amplitude of the SGWB.

Based on the C21 model and the common-process signal in the NANOGrav 12.5-yr dataset, the number of SMBHBs in the local universe is roughly five times higher than that predicted by the M17 model. 
It is evident that there is a significant difference in the number of SMBHBs predicted by these two models. Notably, the number of SMBHB dark sirens significantly influences the constraints on cosmological parameters, especially when the localization of dark sirens is poor~\cite{LIGOScientific:2021aug}. Nevertheless, previous studies have not forecast the capability of SMBHB dark sirens of constraining cosmological parameters based on the C21 model. Consequently, a crucial question arises: Based on the C21 model that is more consistent with SGWB observations, what role can SMBHB dark sirens actually play in measuring cosmological parameters?

To answer this question, we simulate the mock SMBHB dark siren data based on the C21 quasar-based model and the SGWB signal inferred from the NANOGrav 15-yr dataset~\cite{NANOGrav:2023gor}. {It needs to be emphasized that we employ the latest NANOGrav 15-yr dataset rather than the 12.5-yr dataset to ensure that our SMBHB population estimate is consistent with the most up-to-date PTA observations.}
We analyze the ability of SKA-era PTAs to detect the local-universe SMBHBs based on various PTA detection cases, and then combine the mock dark siren data with the 2 Micron All Sky Survey (2MASS) Extended Source Catalog \cite{Jarrett:2000me} to perform cosmological parameter estimations. 
In this work, we focus only on the ability of SMBHB dark sirens to measure $H_0$. This is because our aim is to investigate how SMBHB dark sirens can contribute to resolving the Hubble tension, and we primarily mock SMBHB dark sirens in the local universe, where the distance-redshift relationship is sensitive only to $H_0$. Therefore, we use the SMBHBs dark sirens to infer $H_0$ and fix all other cosmological parameters to the \emph{Planck} 2018 values.

This paper is organized as follows. Section~\ref{sec2} introduces the methodology used in this work. In Section~\ref{sec3}, we report our constraint results and make relevant discussions. The conclusion is given in Section~\ref{sec4}. Unless otherwise stated, we adopt the system of units in which $G = c = 1$ throughout this work.

\section{Methodology}\label{sec2}

\subsection{Subcatalog of 2MASS galaxy catalog}
Massive galaxies are hypothesized to have a higher probability of hosting SMBHBs \cite{Rosado:2013wva}. To select massive galaxies in the local universe, we first estimate their $K$-band absolute magnitudes by $M_{K}=m_{K}-5 \times \log _{10}(d_{\rm L})-25-0.11 \times A_{v}$, where $d_{\rm L}$ represents the luminosity distance to the galaxy. The NANOGrav Collaboration \cite{NANOGrav:2021sdv} considered the impact of peculiar velocities on redshift measurements, allowing for precisely calculating the distances of all galaxies in the 2MASS Redshift Survey (2MRS) catalog \cite{Huchra:2011ii}. The apparent magnitude $m_{K}$ and the extragalactic extinction $A_{v}$ can be obtained from the 2MRS catalog. We implement a $K$-band cutoff with $M_{K} \leq -25$ for selecting galaxies with stellar mass $M_\star \geq 10^{11} M_\odot$. 
Based on this criterion, we select 13172 galaxies from the 2MRS catalog as potential host galaxies for SMBHBs. These candidate host galaxies constitute a new catalog, referred to as the 2MASS catalog in the following text.

\begin{figure}[!htbp]
\centering
\includegraphics[scale=0.5]{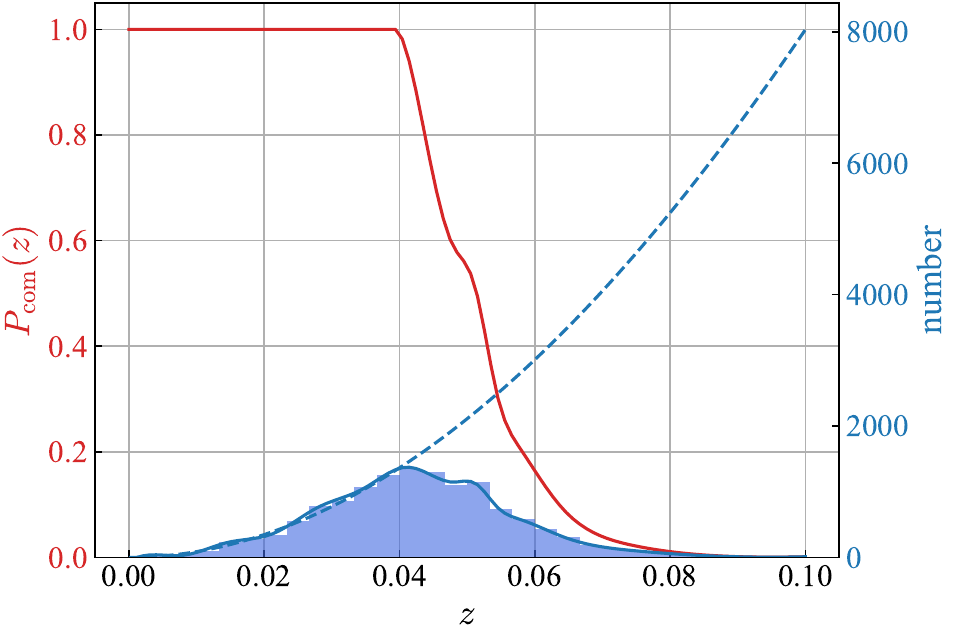}
\caption{\label{pcomplete}The red solid lines represent the subcatalog completeness defined in Eq.~(\ref{pcomplete_equ}) as a function of redshift. The histogram shows the number of galaxies within each redshift bin in the subcatalog. The blue dashed line represents the number of galaxies expected within each redshift bin assuming a uniform distribution of galaxies in comoving volume.}
\end{figure}

Since the 2MASS catalog is incomplete at higher redshifts, we further consider its completeness.
The completeness fraction, $P_{\text{com}}(\mathcal{S})$, can be expressed as
\begin{align}\label{pcomplete_equ}
P_{\text {com}}(\mathcal{S}) \equiv \frac{N_{\text {cat}}(\mathcal{S})}{\bar{n}_{\mathrm{gal}} V_{\rm c}(\mathcal{S})},
\end{align}
where $\mathcal{S} = \mathcal{S}(z, {\hat{\Omega}}; \Delta z, \Delta \Omega)$ represents a volume that is a portion of a cone. This cone is defined by the central redshift $z$ with a range from $z - \Delta z$ to $z + \Delta z$, and the central angle ${\hat{\Omega}}$ with a solid angle $\Delta \Omega$. 
$\bar{n}_{\text{gal}}$ represents the density of galaxies per comoving volume.
$V_{\rm c}$ is the comoving volume of $\mathcal{S}$, and $N_{\text {cat}}$ is the number of galaxies in the catalog within $\mathcal{S}$.
We assume that the completeness of the 2MASS catalog is isotropic and count the number of galaxies in each redshift bin. The redshift range from 0 to 0.1 is equally divided into 30 bins. Additionally, we assume that galaxies are uniformly distributed within the comoving volume and consider the 2MASS catalog complete at $z < 0.04$. Therefore, we can derive the completeness of the 2MASS catalog as a function of redshift using Eq.~(\ref{pcomplete_equ}), which is shown in figure~\ref{pcomplete}.

\subsection{Population models of SMBHBs}
The differential comoving number density of SMBHBs can be described as ${\rm{d}}^3{\Phi_{\mathrm{BHB}}}/{\mathrm{d}} M_{\mathrm{BH}, 1} \mathrm{d} z \mathrm{d} q$, where $M_{\rm BH,1}$ is the mass of the larger SMBHB, and $q$ is the mass ratio defined as $q=M_{\rm BHB}/M_{\rm BH,1}-1$ with $M_{\mathrm{BHB}}$ being the total binary mass. $q$ is assumed to follow a log-normal distribution, $p(q)$, centered at $\log_{10}q = 0$ with a standard deviation of $\sigma_{\log_{10}q} = 0.5~$dex, within the range [0.25, 1.0], similar to the “hiq” model described in Sesana et al. \cite{Sesana:2017lnk}.
While the galactic major merger models can effectively estimate the distribution of SMBHBs, these models are based on the assumptions about the evolution processes of the galaxies and black holes, the galaxy merger rates, and the relationship between the galaxy mass and the black hole mass, etc. However, these assumptions vary greatly among different major merger models, leading to significantly different SMBHB number densities in simulations. It should be noted that the SGWB characteristic amplitude predicted by the M17 model was lower than that of NANOGrav 12.5-yr data. This issue must be addressed if the SGWB signal is considered to be contributed by SMBHBs. 
We note that the C21 model \cite{Casey-Clyde:2021xro} can self-consistently incorporate the SGWB characteristic amplitude as a parameter to infer the distribution of SMBHBs, and the number of nearby SMBHBs inferred from the NANOGrav 12.5-yr data is more than five times the number estimated by the M17 major merger model. Therefore, we choose the C21 model to estimate the distribution of SMBHBs and compare it with the M17 major merger model \cite{Yan:2020ewq,Wang:2022oou}.
\subsubsection{Major merger model}
We convert $M_K$ to the stellar masses for early-type galaxies using the empirical equation, ${\rm log}_{10}M_{\star} = 10.58-0.44(M_{K} + 23)$ \cite{Cappellari:2013sca}. The masses of SMBHBs in these galaxies are determined via the $M_{\rm BHB}$-$M_{\rm bulge}$ relationship from McConnell et al. \cite{McConnell:2012hz}, where the bulge mass, $M_{\rm bulge}$, is approximated to the stellar mass $M_\star$. The mass ratio $q$ of the binary is drawn from a log-normal distribution, $p(q)$, centered at ${\rm log_{10}} q$ = 0 and with the standard deviation of $\sigma_{{\rm log_{10}}q}$ = 0.5 dex, within the range [0.25, 1.0]. Then, the probability of the $j$-th galaxy hosting an SMBHB in the PTA band, $p_j$, is given by
\begin{equation}
p_{j}({M_{\star}}, {z})=\frac{t_{{\rm c},j}}{T_{{\rm life}}} \int_{\mu_\star \geq 0.25} \mathrm{~d} \mu_{\star} \frac{\mathrm{d} N}{\mathrm{~d} t}\left({M_{\star}}, \mu_{\star}, {z}'\right) T_{{\rm life}},
\label{pj}
\end{equation}
where $t_{{\rm c}, j} = (5/256) (\pi f_{\rm low})^{-8/3} (GM_{\rm c}/c^3)^{-5/3}$ is the time to coalescence of the SMBHB in the $j$-th galaxy, with $f_{\rm low}=1$ nHz being the lower limit of the PTA band. $T_{\rm life} = T_{\rm df} + T_{\rm sh}$ is the lifetime of the SMBHB, where $T_{\rm df}$ and $T_{\rm sh}$ are the dynamical friction timescale \cite{BinneyTremaine} and the stellar hardening timescale \cite{Sesana:2015haa}, respectively. 
$\mathrm{d} N / \mathrm{d} t\left(M_{\star}, \mu_{\star}, z'\right)$ represents the galaxy merger rate derived from the Illustris cosmological simulation project \cite{Genel:2014lma,Rodriguez-Gomez:2015aua}, where $M_{\star}$ denotes the stellar mass of the galaxies, $\mu_{\star}$ denotes the progenitor stellar mass ratio, and $z'$ denotes the redshift at which the galaxies merge, calculated at $T_z$ with \emph{Planck} cosmological parameters \cite{Planck:2015fie}.

The number of SMBHBs in the galaxy catalog is $N_{\rm SMBHB} = \sum_{j} p_j$ for one simulation. We perform numerous simulations to obtain a distribution of $N_{\rm SMBHB}$. From this distribution, we select the simulation that yields the $N_{\rm SMBHB}$ value closest to the maximum likelihood. We find that there are 213 sources emitting GWs in the PTA band in the 2MASS catalog. Then, we randomly select 213 galaxies from the 2MASS catalog as the host galaxies of SMBHBs according to the probability $p_j$.

\subsubsection{Quasar-based model}
The characteristic amplitude $h_{\rm c}$ of the GW signal generated by the population of merging SMBHBs can be expressed as \cite{phinney_practical_2001a,Sesana:2008xk,Sesana:2012ak}
\begin{equation}
\begin{aligned}\label{equ3}
h_{\rm c}^{2}(f)=\frac{4}{\pi f^{2}} \iiint & {\rm{d}} M_{\mathrm{BH}, 1} {\rm{d}} z {\rm{d}} q \frac{{\rm{d}}^{3} \Phi_{\mathrm{BHB}}}{{\rm{d}} M_{\mathrm{BH}, 1} {\rm{d}} z {\rm{d}} q} \frac{1}{1+z} \frac{{\rm{d}} E_{\mathrm{gw}}}{{\rm{d}} \ln f_{\rm{r}}},
\end{aligned}
\end{equation}
where $\rm{d}^3\Phi_{\mathrm{BHB}}/d \textit{M}_{\mathrm{BH},1}\,d\textit{z}\,d\textit{q}$ represents the differential comoving number density of SMBHBs, ${\rm d}E_\mathrm{gw}/{\rm d}\ln f_{\rm r}$ is the GW energy emitted per unit logarithmic frequency change, and $f_{\rm r} = f(1 + z)$ is the GW frequency in the rest frame of the source. For a population of SMBHBs in circular orbits, the GW energy per unit logarithmic frequency change is~\cite{thorne_gravitational_1987a,phinney_practical_2001a}
\begin{align}
\frac{{\rm d} E_{\mathrm{gw}}}{{\rm d} \ln f_{\rm{r}}} & = \frac{\left(\pi f_{\rm r}\right)^{2 / 3}}{3} {M_{\rm c}}^{5 / 3},
\end{align}
where the SMBHB chirp mass is given by ${M_{\rm c}}^{5/3} = \frac{q}{{(1+q)}^2}M^{5/3}_{\mathrm{BHB}}$.
Therefore, the differential comoving number density of SMBHBs is the main parameter influencing the value of $h_{\rm c}$. 

Since some quasars could be driven by the mergers of black holes, the populations of quasars and black holes might be correlated. Therefore, the C21 model makes a simplified assumption that the ratio of the differential comoving number density of quasars to that of black holes remains constant:
\begin{align}\label{equ6}
\frac{\rm{d}^{3} \Phi_{\mathrm{BHB}}}{{\rm{d}} M_{\mathrm{BH}, 1} {\rm{d}} z {\rm{d}} q} \propto \frac{\rm{d}^{2} \Phi_{\mathrm{Q}}}{{\rm{d}} M_{\mathrm{BH}} {\rm{d}} z} p(q).
\end{align}
That is to say, the differential comoving number density of SMBHBs is proportional to the differential comoving number density of quasars, ${{\rm{d}}^{2} \Phi_{\mathrm{Q}}}/{{\rm{d}} M_{\mathrm{BH}} {\rm d} z}$, and an assumed mass ratio distribution $p(q)$. The number density of quasars is derived from the quasar luminosity function and Eddington ratio distribution of the lifetime of the quasar from the galaxy merger simulations in Hopkins et al. \cite{Hopkins:2005fb}. 
To estimate the number density of quasars, we convert the differential quasar triggering rate \cite{Hopkins:2006fq}, ${\mathrm{d}^{2} \Phi_{\mathrm{Q}}}/{\mathrm{d} M_{\mathrm{BH}} {\mathrm{d}} t_{\rm r}}$, into a redshift distribution using the $\Lambda$CDM model with WMAP9 data,
\begin{align}\label{equ7}
\frac{\mathrm{d}^{2} \Phi_{\mathrm{Q}}}{\mathrm{d} M_{\mathrm{BH}} \mathrm{d} z}=\frac{\mathrm{d}^{2} \Phi_{\mathrm{Q}}}{\mathrm{d} M_{\mathrm{BH}} \mathrm{d} t_{\rm r}} \frac{\mathrm{d} t_{\rm r}}{\mathrm{d} z}.
\end{align}

According to Eqs.~(\ref{equ6}) and (\ref{equ7}), we can renormalize the SMBHB population with both the SGWB characteristic amplitude and the quasar number density distribution,
\begin{align}\label{equ8}
\frac{\mathrm{d}^{3} \Phi_{\mathrm{BHB}}}{\mathrm{d} M_{\mathrm{BH}, 1} \mathrm{d} z \mathrm{d} q}=\frac{\Phi_{\mathrm{BHB}, 0}}{\Phi_{\mathrm{Q}, 0}} \frac{\mathrm{d}^{2} \Phi_{\mathrm{Q}}}{\mathrm{d} M_{\mathrm{BH}} \mathrm{d} t_{\rm r}} \frac{\mathrm{d} t_{\rm r}}{\mathrm{d} z} \frac{p(q)}{\int_{q \geq 0.25} \mathrm{d} q p(q)},
\end{align}
with

\begin{align}
\Phi_{\mathrm{BHB}, 0} = &\left. \iint_{\substack{M_{\mathrm{BH}, 1} \geq M_{\mathrm{BH}, \mathrm{det}}, \\ q \geq 0.25}} 
\mathrm{d} M_{\mathrm{BH}} \mathrm{d} q 
\frac{\mathrm{d}^{3} \Phi_{\mathrm{BHB}}}{\mathrm{d} M_{\mathrm{BH}, 1} \mathrm{d} z \mathrm{d} q} \right|_{z=0},
\end{align}

\begin{align}\label{equ9}
\Phi_{\mathrm{Q}, 0}=\left.\int_{M_{\mathrm{BH}} \geq M_{\mathrm{BH}, \mathrm{det}}} \mathrm{d} M_{\mathrm{BH}} \frac{\mathrm{d}^{2} \Phi_{\mathrm{Q}}}{\mathrm{d} M_{\mathrm{BH}} \mathrm{d} t_{\rm r}} \frac{\mathrm{d} t_{\rm r}}{\mathrm{d} z}\right|_{z=0}.
\end{align}

{In this model, the characteristic amplitude of the SGWB, $h_c$, is treated as an input parameter. The Bayesian posterior median of the SGWB characteristic amplitude from the NANOGrav 15-yr dataset is $ 2.4 \times 10^{-15} $~\cite{NANOGrav:2023gor}, assuming a spectral index of $13/3$, as expected for SMBHB inspirals. We substitute $h_c=2.4 \times 10^{-15}$ into Eq.~(\ref{equ3}) and use Eqs.~(\ref{equ3})--(\ref{equ9}) to calculate the number of SMBHBs. We estimate that there are approximately 688 SMBHBs in the redshift range (0, 0.1). It should be added that using different values of $h_c$ results in different numbers of SMBHBs but the C21 model is applicable to the observational signals from any PTA project.}


{The possible origins of the current signal observed by PTAs can be broadly classified into astrophysical and cosmological sources. In this study, we adopt an astrophysical interpretation, using $ h_c $ as an input to the C21 model under the assumption that the NANOGrav 15-yr signal is attributed to an ensemble of inspiraling SMBHBs. However, the true origin of the PTA signals remains uncertain. While an incoherent superposition of GWs from numerous SMBHBs provides a natural astrophysical explanation, several alternative cosmological scenarios have been proposed. These include primordial GWs generated by cosmic inflation~\cite{Vagnozzi:2020gtf}, scalar-induced GWs~\cite{Cai:2019bmk,DeLuca:2020agl}, GWs produced during cosmological first-order phase transitions~\cite{Schwaller:2015tja,NANOGrav:2021flc}, GW radiation from cosmic strings~\cite{Ellis:2020ena,Siemens:2006yp} and domain walls~\cite{Ferreira:2022zzo}, among others. Each of these mechanisms could contribute to the GW signal in the nHz frequency range. The current observational data are insufficient to distinguish these potential sources, making future high-precision observations crucial for uncovering the true origin of this signal. }

Considering the completeness of the 2MASS catalog, we use Eq.~(\ref{pcomplete_equ}) to calculate the total completeness fraction of the 688 SMBHBs. This completeness fraction actually refers to the proportion of the host galaxies included in the 2MASS catalog relative to the total host galaxies. We find that it is approximately equal to 490. Therefore, we randomly select 490 galaxies based on their $P_{\rm com}$ values from the 2MASS catalog as SMBHBs’ host galaxies. We assume that the host galaxies of the other 198 SMBHBs are not included in the 2MASS catalog and ignore their contributions to the subsequent calculations for dark sirens.


\begin{figure}[!htbp]
\centering
\includegraphics[width=1\linewidth]{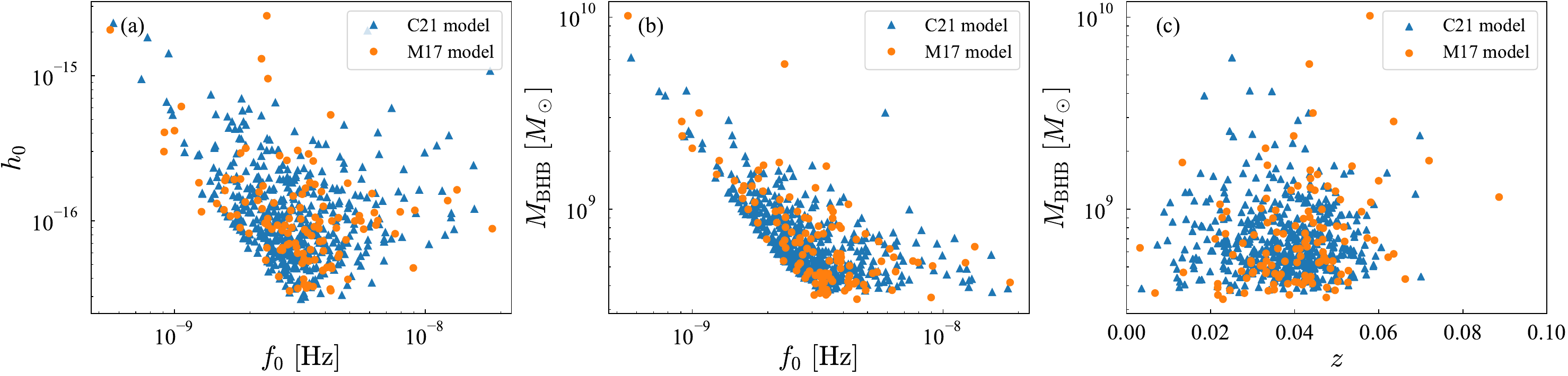}
\caption{\label{h0-f-m-z}Distributions of $h_0$, $f_0$, $M_{\rm BHB}$, and $z$ for simulated SMBHBs. The blue triangles and the orange circles correspond to the C21 and M17 SMBHB population models, respectively.}
\end{figure}

\subsection{GW detection and Fisher analysis}
The timing residuals induced by a single GW source located in the direction represented by the unit vector $\hat{\Omega}$ and measured at time $t$ on Earth can be expressed as
\begin{align}\label{equ1}
s(t, \hat{\Omega})=F^{+}(\hat{\Omega}) \Delta A_{+}(t)+F^{\times}(\hat{\Omega}) \Delta A_{\times}(t).
\end{align}
Here, $F^{+}(\hat{\Omega})$ and $F^{\times}(\hat{\Omega})$ are the antenna pattern functions \cite{Yan:2020ewq}, which can be expressed as
\begin{align}
\begin{aligned}
F^{+}(\hat{\Omega})=&\frac{1}{4(1-\cos \theta)}\left\{\left(1+\sin ^{2} \delta\right) \cos ^{2} \delta_{\mathrm{p}} \cos \left[2\left(\alpha-\alpha_{\mathrm{p}}\right)\right]-\sin 2 \delta \sin 2 \delta_{\mathrm{p}} \cos \left(\alpha-\alpha_{\mathrm{p}}\right)\right. \\
&\left.\quad+\cos ^{2} \delta\left(2-3 \cos ^{2} \delta_{\mathrm{p}}\right)\right\}, \\
F^{\times}(\hat{\Omega})=&\frac{1}{2(1-\cos \theta)}\left\{\cos \delta \sin 2 \delta_{\mathrm{p}} \sin \left(\alpha-\alpha_{\mathrm{p}}\right)-\sin \delta \cos ^{2} \delta_{\mathrm{p}} \sin \left[2\left(\alpha-\alpha_{\mathrm{p}}\right)\right]\right\},
\end{aligned}
\end{align}

where ($\alpha$, $\delta$) and ($\alpha_{\rm p}$, $\delta_{\rm p}$) represent the right ascension and declination of the GW source and MSP, respectively. $\theta$ represents the opening angle between the GW source and MSP,
\begin{align}
\cos \theta=\cos \delta \cos \delta_{\mathrm{p}} \cos \left(\alpha-\alpha_{\mathrm{p}}\right)+\sin \delta \sin \delta_{\mathrm{p}}.
\end{align}
In Eq.~(\ref{equ1}), $\Delta A_{\{+, \times\}} (t)=A_{\{+, \times\}}(t)-A_{\{+, \times\}}\left(t_{\mathrm{p}}\right)$~\cite{Ellis:2012zv} represents the difference between the Earth term $A_{\{+, \times\}}(t)$ and the pulsar term $A_{\{+, \times\}}(t_{\mathrm{p}})$.
$t_{\rm{p}}$ is the time when the GW passes MSP and can be expressed as $t_{\rm{p}}=t-d_{\rm{p}}(1-{\rm{cos}}\theta)/c$, with $d_{\rm{p}}$ representing the pulsar distance.
For evolving SMBHBs in circular orbits, we have
\begin{align}\label{equ2}
\begin{aligned}
A_{+}(t)= \ &\frac{h(t)}{2 \pi f(t)}\{\left(1+\cos ^{2} \iota\right) \cos 2 \psi \sin [\phi(t)]+2 \cos \iota \sin 2 \psi \cos \left[\phi(t)\right]\}, \\
\end{aligned}
\end{align}
\begin{align}
\begin{aligned}
A_{\times}(t)= \ &\frac{h(t)}{2 \pi f(t)}\{\left(1+\cos ^{2} \iota\right) \sin 2 \psi \sin [\phi(t)] -2 \cos \iota \cos 2 \psi \cos \left[\phi(t)\right]\}. \\
\end{aligned}
\end{align}
Here, $\psi$ is the GW polarization angle, $\phi$ is the phase of the GW, $\iota$ is the inclination angle of the GW, and the GW strain amplitude $h(t)$ can be formulated as
\begin{align}
h(t) = 2 \frac{\left(G M_{\mathrm{c}}\right)^{5 / 3}}{c^{4}} \frac{(\pi f(t))^{2 / 3}}{d_{\mathrm{L}}},
\end{align}
where the frequency of GW signals is calculated by the following formula \cite{Peters:1964zz},
\begin{align}\label{f0}
f(t) = \pi^{-1} \left(\frac{G{M}_{\rm c}}{c^3}\right)^{-5/8} \left[\frac{256}{5}(t_{\rm c}-t)\right]^{-3/8}.
\end{align}
Here, $t_{\rm c}$ is uniformly sampled from [100 yr, 26 Myr]. This range represents the time to coalescence of a SMBHB with $M_{\rm BH,1} = 10^9 M_\odot$ and $q = 1$ from 1 nHz to 100 nHz.
$h_0$ and $f_0$ represent the values of $h(t)$ and $f(t)$ when $t=0$, respectively.
Figure~\ref{h0-f-m-z} shows the distribution of mock SMBHBs on the $f_0$-$h_0$, $f_0$-$M_{\rm BHB}$, $z$-$M_{\rm BHB}$ planes.
The inclination angle, the polarization angle, and the initial phase are randomly sampled in the ranges $\cos \iota \in [-1, 1]$, $\psi \in [0, 2\pi]$, and $\phi_0 \in [0, 2\pi]$, respectively.
The GW phase, $\phi(t)$, can be expressed as
\begin{align}
\phi(t)=\phi_0 + \frac{1}{16}\left(\frac{G M_{\mathrm{c}}}{c^{3}}\right)^{-5 / 3}\left\{\left(\pi f_{0}\right)^{-5 / 3}-[\pi f(t)]^{-5 / 3}\right\}.
\end{align}
The square of SNR of a GW signal detected by a PTA is defined as
\begin{align}
\rho^{2}=\sum_{j=1}^{N_{\mathrm{p}}} \sum_{i=1}^{N}\left[\frac{s_{j}\left(t_{i}\right)}{\sigma_{t, j}}\right]^{2},
\end{align}
where $N$ is the total number of the data points of timing residuals for each MSP, $N_{\rm p}$ is the number of MSPs, $s_j(t_i)$ is the timing residual induced by the GW signal in the $j$-th MSP at time $t_i$, and $\sigma_{t,j}$ is the RMS of the white noise of the $j$-th MSP. The SNR threshold is set to 10 in the simulation.

As this paper focuses on the impact of SMBHB population models on dark sirens, we consider only white noise as the noise source, without taking into account red noise or SGWBs. According to the analyses in other papers, taking into account red noise and SGWB could reduce SNRs of GW signals \cite{Liu:2023kwn}, and red noise could worsen the measurement precision of cosmological parameters by 3--5 times \cite{Wang:2022oou}. Therefore, we can roughly estimate that, red noise and SGWB would also worsen the constraint precision of $H_0$ presented in this paper by a few times. A more detailed analysis of red noise and SGWB will be addressed in our future works.

We utilize the Fisher information matrix for parameter estimations. For a PTA comprising $N_{\rm p}$ independent MSPs, the Fisher information matrix can be expressed as
\begin{align}
\Gamma_{a b}=\sum_{j=1}^{N_{\rm p}} \sum_{i=1}^{N} \frac{\partial s\left(t_{i}\right)}{\sigma_{t, j} \partial \theta_{a}} \frac{\partial s\left(t_{i}\right)}{\sigma_{t, j} \partial \theta_{b}},
\end{align}
where $\boldsymbol{\theta}$ denotes the free parameters to be estimated.

For each GW source, the timing residuals depend on $N_{\rm {p}}+8$  system parameters, including eight parameters of the GW source (i.e., $M_{\rm c}, \alpha,\delta, \iota, \psi, \phi_0, f_0, d_{\rm L}$) and the distances of MSPs $d_{\rm{p},\textit{j}}$ $( j = 1, 2,\cdots, N_{\rm p})$. With additional EM information priors, the estimation precision of the parameter can be improved by adding to the appropriate diagonal element of the Fisher information matrix, i.e., 
\begin{align}
\label{prior}
\Gamma_{a b} \rightarrow \Gamma_{a b}-\left\langle\frac{\partial^{2} \ln P\left(\theta_{i}\right)}{\partial \theta_{a} \partial \theta_{b}}\right\rangle,
\end{align}
where $P(\theta_i)$ is the prior distribution of the parameter $\theta_i$~\cite{Albrecht:2009ct}. Since the disk direction is randomly distributed in a 4$\pi$ solid angle, the prior distribution of the inclination angle $\iota$ is given by $P(\iota) \propto \sin (\iota)$. Consequently, we can modify the Fisher information matrix element $\Gamma_{i i}$ to $\Gamma_{i i} \rightarrow \Gamma_{i i}+\frac{1}{\sin ^{2} \iota}$ with $\theta_i = \iota$.  

The 1$\sigma$ absolute error of the parameter $\theta_a$ can be estimated by 
\begin{align}
\Delta \theta_{a}=\left(\Gamma^{-1}\right)_{a a}^{1 / 2}.
\label{fishererror}
\end{align}
The angular resolution $\Delta \Omega$ can be calculated according to the uncertainties of the parameters,
\begin{equation}
\begin{aligned}
\Delta \Omega=2 \pi \sqrt{(\cos \delta \Delta \delta \Delta \alpha)^{2}-\left(\cos \delta \frac{\left(\Gamma^{-1}\right)_{\delta \alpha}}{\Delta \delta \Delta \alpha}\right)^{2}}.
\end{aligned}
\end{equation}
For the parameter $d_{\rm L}$, we first estimate its instrumental error by Eq.~(\ref{fishererror}). Besides, we also consider the error arising from weak lensing, which is expressed by the fitting formula \cite{Hirata:2010ba,Tamanini:2016zlh},
\begin{align}
\Delta{d_{\mathrm{L}}^{\text {lens}}}(z) & = d_{\mathrm{L}}(z) \times 0.066\left[\frac{1-(1+z)^{-0.25}}{0.25}\right]^{1.8}.
\end{align}
The total error of $d_{\rm L}$ is expressed as
\begin{align}
\Delta{d_{\mathrm{L}}} & = \sqrt{\left(\Delta{d_{\mathrm{L}}^{\rm inst}}\right)^{2}+\left(\Delta{d_{\mathrm{L}}^{\rm lens}}\right)^{2}}.
\end{align}

\begin{figure}[!htbp]
\centering
\includegraphics[scale=0.7]{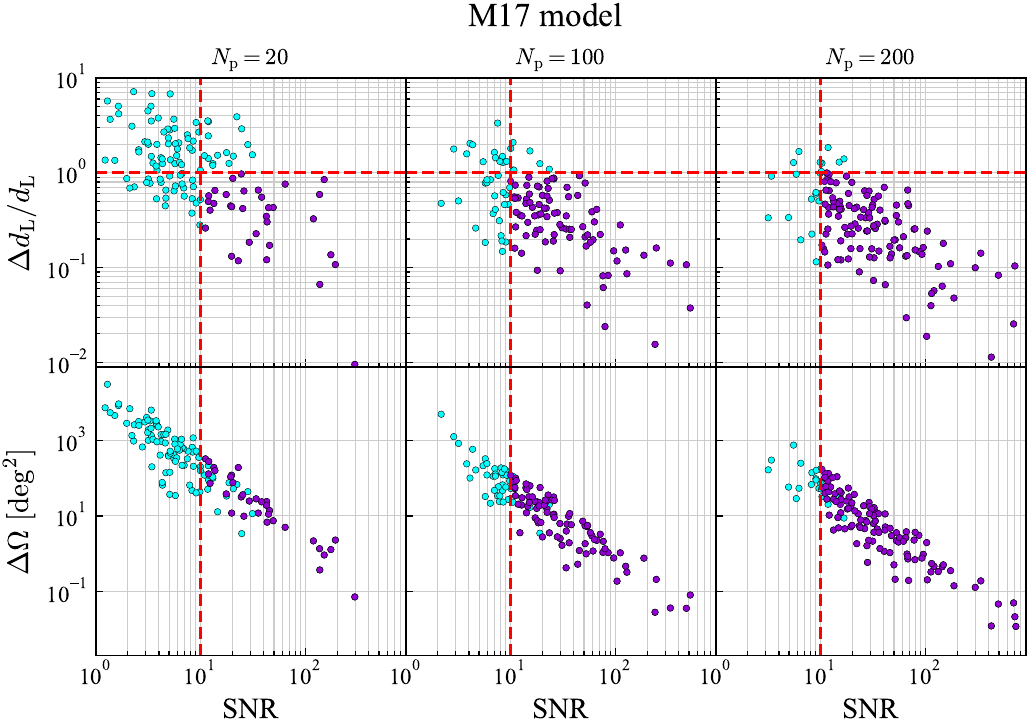}
\includegraphics[scale=0.7]{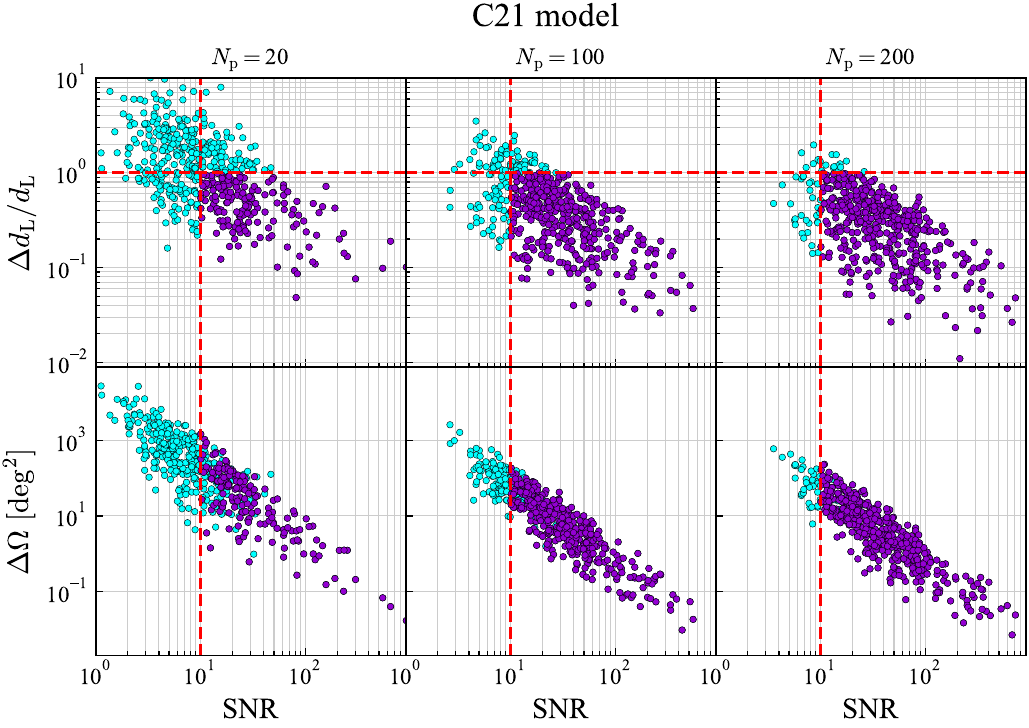}
\caption{{\label{location}} Distributions of the $d_{\rm L}$ relative errors and the angular resolutions of simulated SMBHBs versus SNR for different PTA detection cases. The SMBHBs are simulated based on the M17 model (upper panel) and the C21 model (lower panel), respectively. 
From left to right, the three columns correspond to different PTA detection cases with $N_{\rm p}$ set to 20, 100, and 200, respectively. Observations are conducted over ten years with $\sigma_t = 20$ ns. The red dotted vertical and horizontal lines denote the thresholds of $\rho=10$ and $\Delta{d_{\rm L}}/d_{\rm L} = 1$, respectively. The purple dots represent the GW events with SNR $>$ 10 and $\Delta{d_{\rm L}}/d_{\rm L} < 1$, which are detectable and suitable as dark sirens.}
\end{figure}

The distances of MSPs are expected to be measured via the timing parallax method in the SKA era. We estimate the uncertainties of MSPs' distances by using Eq.~(18) in Lee et al. \cite{Lee:2011et},
\begin{align}
\sigma_{d_{\rm p}} \simeq \frac{2.34}{\cos ^{2} \beta_{\rm p}}\left(\frac{N}{100}\right)^{-1/2}\left(\frac{d_{\rm p}}{1~\mathrm{kpc}}\right)^{2}\left(\frac{\sigma_{t}}{10~\mathrm{ns}}\right)~\mathrm{pc},
\end{align}
where $\beta_{\rm p}$ represents the ecliptic latitudes of MSPs. In the Fisher matrix, we take into account the priors of MSPs' distances {using} Eq.~(\ref{prior}). In this context, $P(d_{\rm p})$ is a Gaussian distribution with mean $d_{\rm p}$ and the standard deviation $\sigma_{d_{\rm p}}$. 
Under this assumption, the derivatives of GW frequencies can be obtained, and the chirp masses of the SMBHBs can be further measured. This enables us to break the degeneracy between ${M_{\rm c}}$ and $d_{\rm L}$, and therefore precisely infer $d_{\rm L}$ from GW amplitudes.
The distributions of the $d_{\rm L}$ relative errors and the angular resolutions of simulated SMBHBs versus SNR based on 20, 100, and 200 MSPs with $\sigma_t$ = 20 $\rm ns$ and 10-year observation are displayed in figure~\ref{location}. The sky locations of MSPs are randomly selected from the Australia Telescope National Facility pulsar catalog \cite{Manchester:2004bp}.


\subsection{Inference of the Hubble constant}
In this work, we adopt the flat $\Lambda$CDM model to make the cosmological analysis. The Hubble parameter, which quantifies the rate of expansion of the universe at redshift 
$z$, can be expressed as
\begin{equation}
H(z) = H_{0} \sqrt{\Omega_{\rm m}(1+z)^{3}+1-\Omega_{\rm m}},
\end{equation}
where $H_0$ is the Hubble constant and $\Omega_{\rm m}$ is the current matter density parameter. The luminosity distance at redshift $z$ is given by
\begin{equation}\label{dl2z}
d_{\rm L}(z) = c(1+z) \int_{0}^{z} \frac{1}{H(z^{\prime})} \mathrm{d} z^{\prime},
\end{equation}
where $c$ is the speed of light in vacuum. Throughout this work, 
we adopt the flat $\Lambda$CDM model to generate the dark siren data with the best-fit values of $\Omega_{\rm m}=0.3111$ and $H_0=67.66 \rm~km~s^{-1}~Mpc^{-1}$ given by the \emph{Planck} 2018 TT, TE, EE+lowE+lensing+BAO data \cite{Planck:2018vyg}. Based on the flat $\Lambda$CDM model, we utilize the GW signals from SMBHBs as dark sirens to infer $H_0$, fixing the other cosmological parameters to the \emph{Planck} 2018 results.

Using the Bayesian method to infer the cosmological parameters, the posterior probability distribution of $H_0$ can be expressed as
\begin{equation}{\label{equ_bay}}
\begin{aligned}
&p\left(H_{0} \mid \mathcal{D}_{\mathrm{GW}}, \mathcal{D}_{\mathrm{EM}}\right)= \frac{\prod_{i=1}^{N_{\mathrm{GW}}} p\left(\mathcal{D}_{\mathrm{GW}, i}, \mathcal{D}_{\mathrm{EM}, i} \mid {H_0}\right)p({H_0})}{\gamma(H_0)},
\end{aligned}
\end{equation}
where $\mathcal{D}_{\mathrm{GW}}$ and $\mathcal{D}_{\mathrm{EM}}$ represent the detected GW data and the EM data, respectively. $p({H_0})$ is the prior distribution of $H_0$, and $N_{\mathrm{GW}}$ is the number of GW events. $p\left(\mathcal{D}_{{\mathrm{GW},i}}, \mathcal{D}_{{\mathrm{EM},i}}\mid {H_0}\right)$ is the likelihood function of a single GW event, which can be written as
\begin{equation}\label{equ20}
\begin{aligned}
&p\left(\mathcal{D}_{\mathrm{GW}, i}, \mathcal{D}_{\mathrm{EM}, i} \mid H_{0}\right) \\
&= \int p\left(\mathcal{D}_{\mathrm{GW}, i}, \mathcal{D}_{\mathrm{EM}, i}, {d}_{\rm{L}}, \alpha, \delta, z \mid H_{0}\right) \mathrm{d} {d}_{\rm{L}} \mathrm{d} \alpha \mathrm{d} \delta \mathrm{d} z \\
&= \int p\left(\mathcal{D}_{\mathrm{GW}, i} \mid {d}_{\rm{L}}, \alpha, \delta\right) p\left(\mathcal{D}_{\mathrm{EM}, i} \mid z, \alpha, \delta\right) p\left({d}_{\rm{L}} \mid z, H_{0}\right)p_{0}\left(z, \alpha, \delta \mid H_{0}\right) \mathrm{d} {d}_{\rm{L}} \mathrm{d} \alpha \mathrm{d} \delta \mathrm{d} z \\
&= \int p\left(\mathcal{D}_{\mathrm{GW}, i} \mid {d}_{\rm{L}}, \alpha, \delta\right) p\left(\mathcal{D}_{\mathrm{EM}, i} \mid z, \alpha, \delta\right)\delta\left.(d_{\rm{L}}-\hat{d}_{\rm{L}}\left(z, H_{0}\right)\right.) p_{0}\left(z, \alpha, \delta \mid H_{0}\right) \mathrm{d} d_{\rm{L}} \mathrm{d} \alpha \mathrm{d} \delta \mathrm{d} z \\
&= \int p\left.(\mathcal{D}_{\mathrm{GW}, i} \mid \hat{d}_{\rm L}\left(z, H_{0}\right), \alpha, \delta\right.) p\left(\mathcal{D}_{\mathrm{EM}, i} \mid z, \alpha, \delta\right) p_{0}\left(z, \alpha, \delta \mid H_{0}\right) \mathrm{d} \alpha \mathrm{d} \delta \mathrm{d} z.
\end{aligned}
\end{equation}

\begin{figure}[!h]
\centering
\includegraphics[scale=0.5]{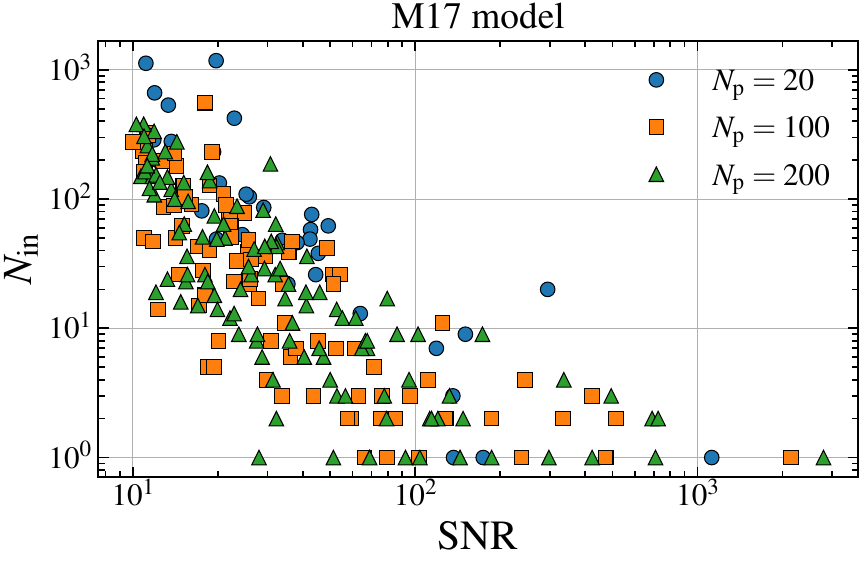}
\includegraphics[scale=0.5]{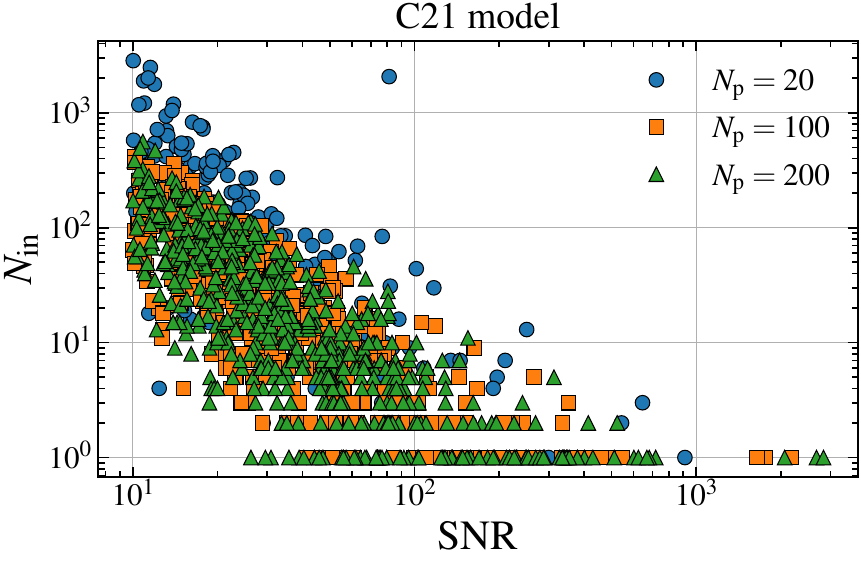}
\caption{\label{Nin}Correlations between SNR and the number of potential host galaxies $N_{\rm in}$ for the mock GW events based on the M17 model (upper panel) and the C21 model (lower panel) under different PTA detection cases. The dots with different colors and shapes represent different PTA detection cases with $N_{\rm p}=20, 100,\rm{and}~200$, respectively. The observation period is 10 years and $\sigma_t$ is set to 20 ns.}
\end{figure}

$p\left.(\mathcal{D}_{\mathrm{GW}, i} \mid \hat{d}_{\mathrm{L}}\left(z, H_{0}\right), \alpha, \delta\right.)$ is the likelihood of the GW data, which can be expressed as
\begin{align}
p\left.(\mathcal{D}_{\mathrm{GW}, i} \mid \hat{d}_{\mathrm{L}}\left(z, H_{0}\right), \alpha, \delta\right.) \propto e^{-\chi^{2} / 2},
\end{align}
with $\chi^2 = (x-x_{\rm gw})^T\mathcal{C}^{-1}(x-x_{\rm gw})$, where $\mathcal C$ is the $3 \times 3$ covariance matrix only relevant to $(d_{\rm{L}}, \alpha, \delta)$, obtained from the Fisher information matrix. $x = (\hat{d}_{\rm{L}}(z,H_0), \alpha, \delta)$ represents the three-dimensional (3D) position in the sky.
$\boldsymbol{x}_{\rm gw}=(d_{{\rm L},i}, \alpha_{ i}, \delta_{i})$ represents the 3D position of the GW source. 
Considering the bias between the true value and the posterior median from the actual observation, we use the Fisher matrix to determine the Gaussian distribution of $ x_{\rm gw} $ and randomly sample $ x_{\rm gw} $ from this distribution.
We establish the boundary of the GW source’s localization volume based on the 3D measurement errors of $x_{\rm gw}$. The galaxies whose positions satisfying $\chi^2 \leq 11.34$ (corresponding to $99\%$ confidence) can be considered as the potential host galaxies of the GW source. The number of potential host galaxies for a GW source is defined as $N_{\rm in}$, which is dependent on the prior values of $H_0$. We display $N_{\rm in}$ as a function of SNR in figure~\ref{Nin}, fixing $H_0=67.66~\rm km\ s^{-1}~\rm Mpc^{-1}$. Due to the lack of EM signals in dark sirens, $p\left(\mathcal{D}_{\mathrm{EM},i} \mid z, \alpha, \delta\right)$ can be set to a constant value~\cite{Chen:2017rfc}.

For the potential candidate host galaxies, $p_0(z, \alpha, \delta \mid H_0)$ in Eq.~(\ref{equ20}) can be expressed as
\begin{equation}
\begin{aligned}
p_{0}\left(z, \alpha, \delta \mid H_{0}\right) = & f_{\rm com} p_{\text {cat}}(z, \alpha, \delta)+(1-f_{\rm com})p_{\text {miss}}\left(z, \alpha, \delta \mid H_{0}\right),
\end{aligned}
\end{equation}
where $f_{\rm com}$ is the completeness fraction of the catalog, which can be expressed as
\begin{equation}
f_{\rm com} = \frac{N_{\text {cat}}}{\bar{n}_{\mathrm{gal}} V_{\rm max}},
\end{equation}
where $N_{\text {cat}}$ is the number of galaxies in the galaxy catalog, and $V_{\rm max}$ is the entire comoving volume bounded by the upper limit of redshift, i.e., $z_{\rm max}= 0.1$. $p_{\text {cat}}(z, \alpha, \delta)$ is the observed galaxy distribution function provided by the galaxy catalog, which is expressed as 
\begin{equation}
\begin{aligned}
p_{\text {cat}}(z, \alpha, \delta)=\frac{1}{N_{\text {in}}} \sum_{j=1}^{N_{\text {in}}} &\mathcal{N}\left(z_{j}, \sigma_{z, j}\right) \delta\left(\alpha-\alpha_{j}\right)\delta\left(\delta-\delta_{j}\right),
\end{aligned}
\end{equation}
where $\mathcal{N}$ represents the Gaussian distribution with mean $z_j$ and the standard deviation $\sigma_{z,j} = (1+z) {\sqrt {\langle{v_{\rm p}}^2\rangle}}/{c}$~\cite{Hogg:1999ad, Muttoni:2023prw}. 
${\sqrt {\langle{v_{\rm p}}^2\rangle}}$ is the standard deviation of the radial peculiar velocity of a galaxy, set to $500~\rm km~s^{-1}$~\cite{Henriques:2011xn}.   
The expression for $p_{\rm miss}$ is given by
\begin{equation}
\begin{aligned}
p_{\text{miss}}(z, \alpha, \delta \mid H_{0}) \propto (1 - p_{\text{com}}(z, \alpha, \delta)) \frac{\mathrm{d} V_{\rm c}}{\mathrm{d} z \mathrm{d} \alpha \mathrm{d} \delta}.
\end{aligned}
\end{equation}

The normalization term $\gamma(H_0)$ in Eq.~(\ref{equ_bay}) can be written as 
\begin{equation}
\begin{aligned}
\gamma\left(H_{0}\right)= \int &p_{\operatorname{det}}^{\mathrm{GW}}\left({d}_{\mathrm{L}}\left(z, H_{0}\right), \alpha, \delta\right) p_{\operatorname{det}}^{\mathrm{EM}}(z, \alpha, \delta)p_{0}(z, \alpha, \delta) \mathrm{d} z \mathrm{d} \alpha \mathrm{d} \delta,
\end{aligned}
\end{equation}
where $p_{\mathrm{det}}^{\mathrm{GW}}$ and $p_{\mathrm{det}}^{\mathrm{EM}}$ represent the GW detection probability and the EM detection probability, respectively. The GW detection probability can be expressed as
\begin{align}
\begin{aligned}
&p_{\operatorname{det}}^{\mathrm{GW}}\left(d_{\mathrm{L}}\left(z, H_{0}\right), \alpha, \delta\right)=\int_{\mathcal{D}_{\mathrm{GW}}>\mathcal{D}_{\mathrm{GW}}^{\text {th}}} p\left(\mathcal{D}_{\mathrm{GW}} \mid {d}_{\mathrm{L}}\left(z, H_{0}\right), \alpha, \delta\right) \mathrm{d} \mathcal{D}_{\mathrm{GW}},
\end{aligned}
\end{align}
where $\mathcal{D}^{\rm th}_{\rm GW}$ denotes the SNR threshold for GW detections and is set to 10. The EM detection probability can be expressed as $p_{\mathrm{det}}^{\mathrm{EM}}(z, \alpha, \delta) \propto \mathcal{H}\left(z_{\max }-z\right)$, where $\mathcal{H}$ is the Heaviside step function and $z_{\max}=0.1$.

\begin{figure}[!htbp]
\centering
\includegraphics[scale=0.3]{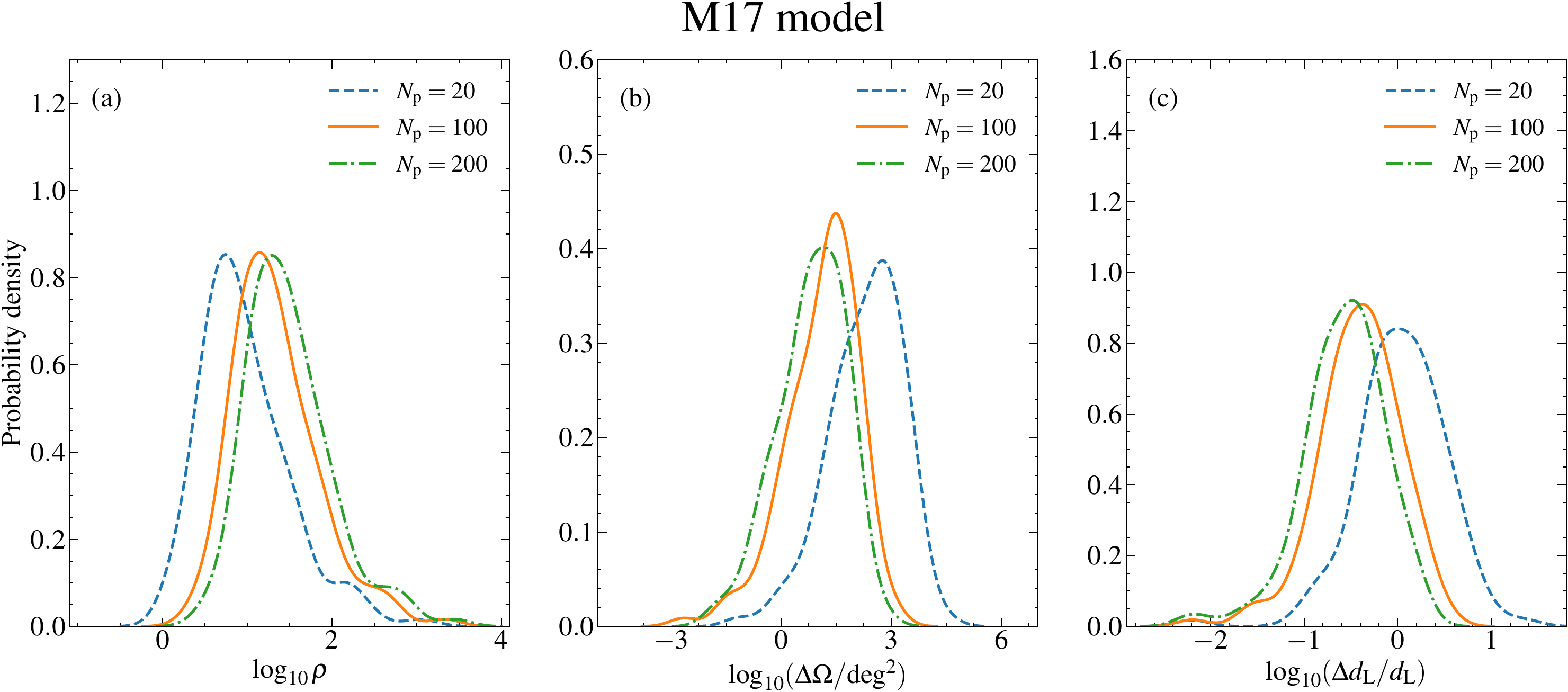}
\includegraphics[scale=0.3]{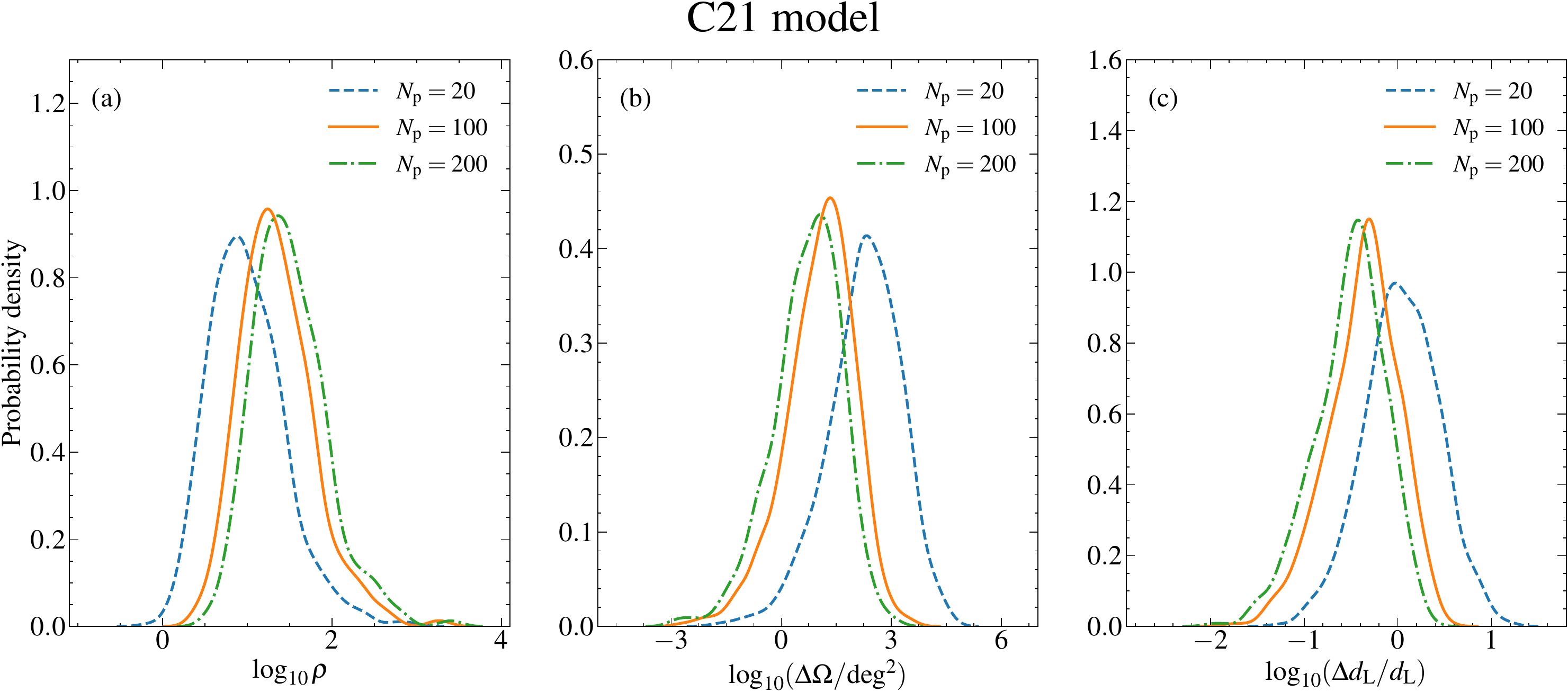}\caption{\label{kdeplot}Distributions of $\rho$, $\Delta\Omega$, and $\Delta d_{\rm L}/d_{\rm L}$ for simulated SMBHBs in a 10-year observation with $\sigma_t$ = 20 ns. The upper and lower panels correspond to the M17 and C21 models, respectively. The lines in different colors and line styles represent different PTA detection cases with $N_{\rm p}=20, 100, \rm{and}~200$ respectively.}
\end{figure}

\begin{table*}[!htbp]
\centering
\normalsize
\setlength\tabcolsep{12pt}
\renewcommand\arraystretch{1}
\begin{tabular}{cccccccccccccccccccc}
\hline \hline
\multirow{2}{*}{Detection case} & \multirow{2}{*}{$T_{\text{span}}~[\rm yr]$} & \multicolumn{2}{c}{M17 model} & & \multicolumn{2}{c}{C21 model} \\ \cline{3-4} \cline{6-7}
& &{$N_{\text{s}}$} & $\varepsilon(H_0)$ & & $N_{\text{s}}$ & $\varepsilon(H_0)$\\ \hline
\multirow{4}{*}{$\sigma_{t}=20\ [\text{ns}], N_{\rm p}=20$} 
& 5.0  & 49 & 7.75\% & & 174 & 3.56\% \\
& 10.0 & 63 & 4.49\% & & 237 & 2.91\% \\
& 15.0 & 67 & 3.10\% & & 274 & 2.27\% \\
& 20.0 & 80 & 2.58\% & & 295 & 1.82\% \\ \hline
\multirow{4}{*}{$\sigma_{t}=20\ [\text{ns}], N_{\rm p}=100$} 
& 5.0  & 105 & 2.21\% & & 377 & 1.48\% \\
& 10.0 & 130 & 1.78\% & & 421 & 1.18\% \\
& 15.0 & 138 & 1.54\% & & 451 & 0.97\% \\
& 20.0 & 150 & 1.44\% & & 461 & 0.91\% \\ \hline
\multirow{4}{*}{$\sigma_{t}=20\ [\text{ns}], N_{\rm p}=200$} 
& 5.0  & 133 & 1.72\% & & 439 & 1.11\% \\
& 10.0 & 152 & 1.36\% & & 457 & 0.88\% \\
& 15.0 & 164 & 1.18\% & & 481 & 0.77\% \\
& 20.0 & 173 & 1.11\% & & 487 & 0.68\% \\ 
\hline \hline
\end{tabular}
\caption{\label{tab1}Relative errors of $H_0$ using the mock dark sirens from different PTA detection cases based on the M17 model and the C21 model, respectively. Here, $N_{\rm s}$ represents the number of detected SMBHBs with $\rho>10$ and $\Delta{d_{\rm L}}/d_{\rm L} < 1$ for different $T_{\rm span}$ and $N_{\rm p}$.}
\end{table*}

\section{Results and discussions}\label{sec3}

\begin{figure}[!h]
\centering
\includegraphics[scale=0.6]{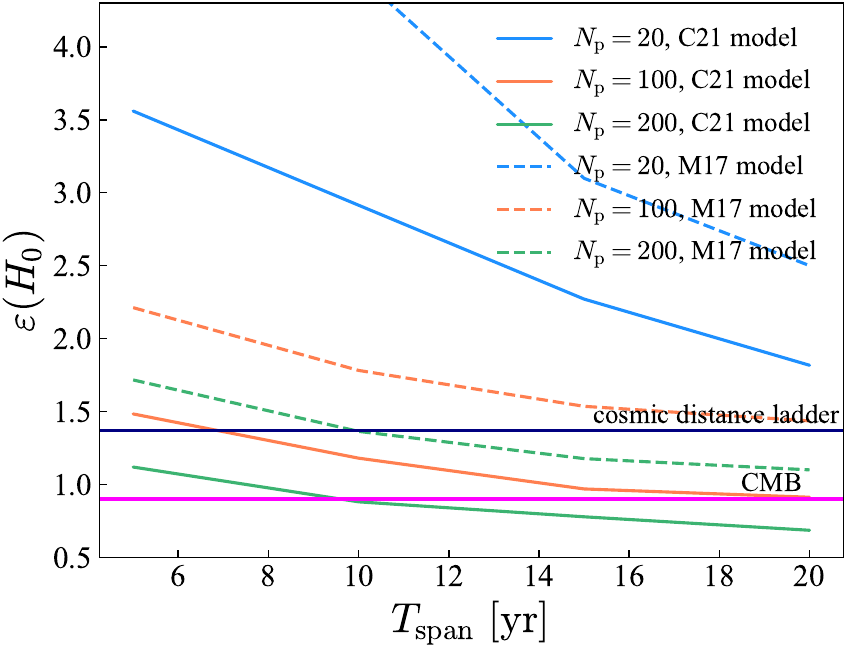}\ \hspace{0.4cm}
\caption{\label{result}The constraint precision of $H_0$ from the mock dark sirens as a function of the PTA observation time span. The dotted and solid lines represent the constraint results from the M17 and C21 models, respectively. Different colors represent $N_{\rm p}=20, 100, \rm{and}~200$ respectively. Here, $\sigma_t$ is set to 20 ns.}
\end{figure}

In this section, we compare the two SMBHB population models, discuss the GW parameter estimations, and present the constraint results for $H_0$ from various PTA detection cases.
As shown in figure~\ref{h0-f-m-z}, the $h_0$, $z$, $M_{\rm BHB}$, and $f_0$ distributions of SMBHBs simulated from the two population models are essentially similar. Nevertheless, the numbers of simulated SMBHBs for the two population models are 490 and 213, respectively, showing that the C21 model has a noticeably larger number of SMBHBs than the M17 model. We consider the 10-year observations based on the C21 model as the representative example of PTA observations for the following discussions.



In figure~\ref{kdeplot}, we present the distributions of $\rho$, $\Delta\Omega$, and $\Delta d_{\rm L}/d_{\rm L}$ based on the 20, 100, and 200 MSPs with $\sigma_t = 20$ ns. Note that the calculations are based on 213 (M17 model) and 490 (C21 model) simulated SMBHBs at $z < 0.1$.
As shown in the left panel of figure~\ref{kdeplot}, fewer available MSPs result in lower SNRs, thus showing worse localization abilities. We note that the localization capability with 100 MSPs is much better than that of 20 MSPs, but 200 MSPs are slightly better than 100 MSPs. This indicates that 100 MSPs seem to be sufficient and more economical, as further increasing the number of MSPs does not significantly enhance the localization ability.

For the mock SMBHBs from the two population models, we consider PTA cases of monitoring 20, 100, and 200 MSPs with $\sigma_t = $ 20 ns to detect them. The observation time spans are $T_{\rm span} = 5, 10, 15, 20$ years, respectively, with a cadence of 2 weeks that is the typical cadence used in current PTAs. The detected SMBHBs are used as dark sirens to constrain $H_0$. 

Figure~\ref{result} illustrates the relative errors for $H_0$ as a function of $T_{\rm span}$. Table~\ref{tab1} shows the relative error of $H_0$, defined as $\varepsilon(H_0)=\sigma(H_0)/H_0$. We observe that the PTAs with $N_{\rm p}=20$ give the poorest measurement precision for $H_0$. Based on the C21 model, the constraint precision of $H_0$ is 3.56\% for a 5-year observation, improving to 1.82\% over a 20-year observation. These values are worse than the constraint precision obtained from current distance ladder measurements. Increasing the number of MSPs to 100 significantly enhances the precision of $H_0$ compared to using 20 MSPs. With a 5-year observation, the constraint precision of $H_0$ is 1.48\% and can be improved to 0.91\% with a 20-year observation; both values are close to or better than 1\%. Furthermore, monitoring 200 MSPs provides even better constraints on $H_0$, achieving precisions ranging from 1.11\% to 0.68\% over 5 to 20 years of observations. {Considering that SKA is expected to begin operations around 2030, our optimistic estimate ($\sigma_t = 20$ ns, $N_{\rm p} = 200$) suggests that PTA dark sirens can achieve a measurement precision of approximately 1\% for $H_0$ over a 10-year observation, which would be around 2040. Nevertheless, under a more conservative noise assumption and considering the impact of the SGWB, the first results are more likely to be obtained after 2050. In figure \ref{result}, we plot the current $H_0$ values obtained from the distance ladder measurements and the \emph{Planck} CMB data. It should be noted that when the precision of the PTA dark siren method in measuring $H_0$ reaches its expected level in the future, the measurement accuracy of the distance ladder method will be also significantly enhanced. Although it is difficult for the PTA dark siren method to surpass the distance ladder method in terms of precisely measuring $H_0$, it can still provide assistance in resolving the Hubble tension as an independent late-universe probe.}

{It is worth noting that our results are influenced by the value of $h_c$. Since $ h_c^2 \propto \Phi_{\rm BHB,0} $, a larger $ h_c $ corresponds to a greater inferred number of SMBHBs. This increases the number of detectable SMBHBs serving as dark sirens, and further enhances the measurement precision of $H_0$. Compared with the NANOGrav 12.5-yr data, the $h_c$ value inferred from the NANOGrav 15-yr dataset suggests a larger SMBHB population. However, the $h_c$ values reported by NANOGrav \cite{NANOGrav:2023gor}, EPTA+InPTA \cite{EPTA:2023fyk}, PPTA \cite{Zic:2023gta} and CPTA \cite{Xu:2023wog} in 2023 are generally consistent. Therefore, using different $h_c$ values provided by various PTA projects will not alter our main conclusions.}

\begin{figure*}[!b]
\centering
\includegraphics[scale=0.7]{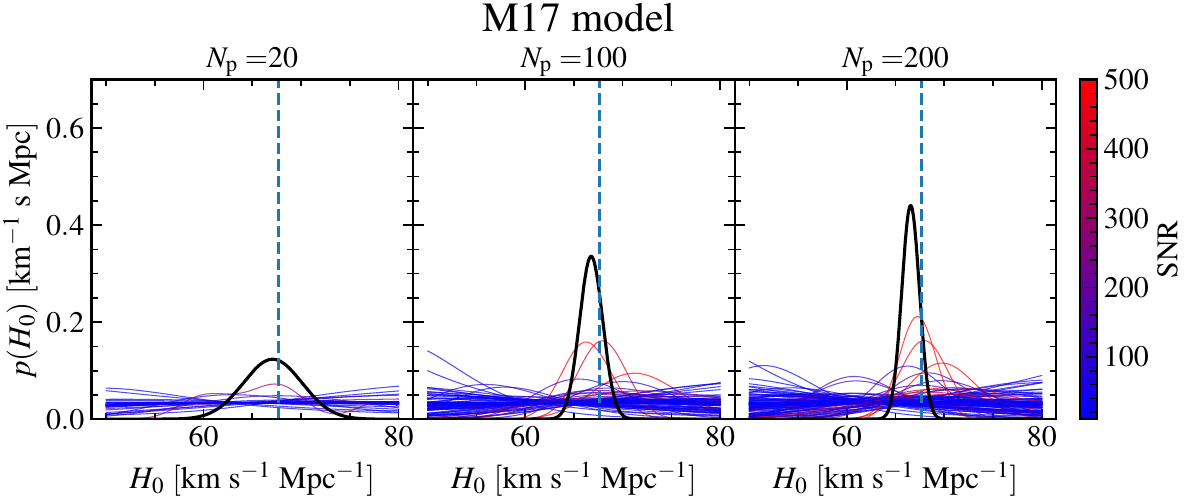}
\includegraphics[scale=0.7]{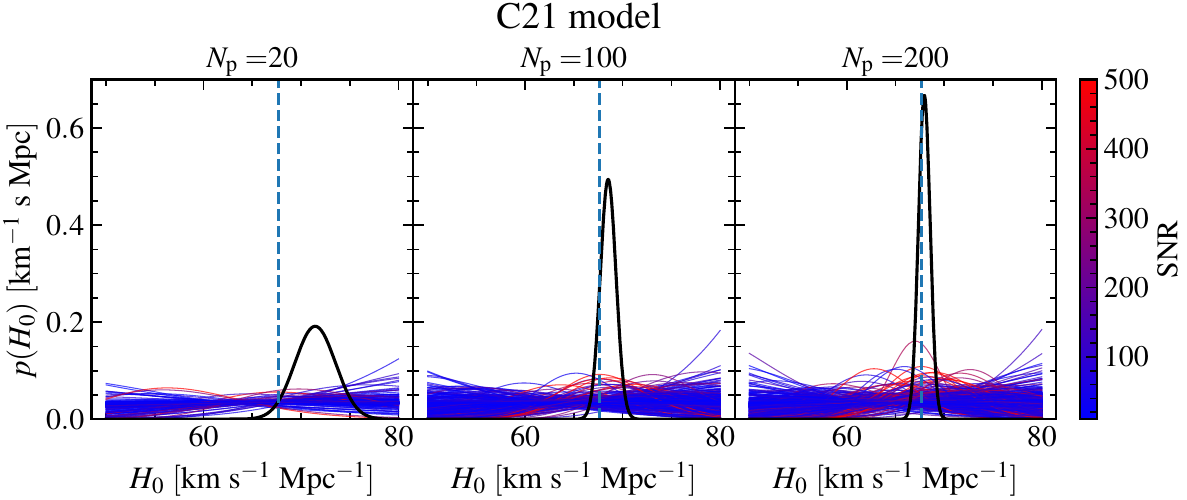}\caption{The marginalized posterior of $H_0$ using the mock dark sirens data based on the M17 model (upper panel) and the C21 model (lower panel) in different PTA detection cases. In each panel, PTA detection cases are considered with $N_{\rm p}$ of 20, 100, and 200, observed over ten years with $\sigma_t$ = 20 ns. The colored lines show the posterior probability density of $H_0$ for each source with colors representing their SNRs, and the thick black line shows the combined posterior probability density of $H_0$. The fiducial value of $H_0$ is shown as a vertical dashed line.}
\label{H0post}
\end{figure*}

Figure~\ref{H0post} presents the constraints on $H_0$ obtained from individual GW events as well as the combined results. The colored lines represent the constraints from individual events and the black line represents the combined result of all events.
There is a notable discrepancy between the two SMBHB population models in terms of the $H_0$ posterior distributions given by the individual events. In the M17 model, there are several events whose SNRs far exceed the others, but in the C21 model, there is no significant difference in the SNRs given by all the events. This discrepancy can also be seen in figure~\ref{h0-f-m-z}: Several SMBHBs (represented by the orange circles) in the M17 model have significantly higher masses and GW strains than the others. This is due to the fact that the mechanisms of the two models in predicting the SMBHB populations are substantially different.
The M17 model predicts the masses of SMBHBs based on the apparent magnitude of the galaxies in the galaxy catalog. 
A few of these galaxies have significantly larger masses, and the SMBHBs inhabiting them could produce the GW
signals with notably higher SNRs than the other SMBHBs. However, for the C21 model, its SMBHB distribution is derived from the SGWB signal, which is the overall superposition of many GW events and does not reflect the information of individual sources. As a result, the SMBHB distribution predicted by the C21 model is statistical rather than individualistic. Therefore, in the C21 model, no event exhibits a SNR that is significantly higher than other events.

{PTA dark sirens primarily rely on low-redshift ($z < 0.1$) GW sources and galaxy catalogs. Within this range, the distance-redshift relation can be approximated by Hubble's law and has limited dependence on late-time cosmological extended models with new physics, such as time-varying dark energy~\cite{Chevallier:2000qy,Linder:2002et}, interacting dark energy~\cite{Amendola:1999er,Wang:2021kxc}, phantom dark energy~\cite{Brown:2004cs,Zhang:2007yu}, decaying dark matter~\cite{Ibarra:2008jk,He:2019bcr}, and spatial curvature~\cite{DiValentino:2019qzk,Qi:2020rmm}. Consequently, these additional parameters cannot be effectively constrained by the low-redshift PTA dark sirens. If these extended models are considered, the additional introduced parameters would exhibit degeneracy with $H_0$, thereby slightly diminishing the measurement precision of $H_0$.}

Compared with the dark siren analysis in Wang et al. \cite{Wang:2022oou}, under the same detection conditions and population models, the constraint precision of $H_0$ is improved in this paper. This is mainly due to the fact that we simulate SMBHBs over a broader range of redshift ($0<z<0.1$), resulting in a larger number of GW sources available as dark sirens. We further consider the effect of incompleteness for galaxies at $z>0.04$.
Wang et al. \cite{Wang:2022oou} also indicated some advantages of nHz GW standard sirens over the standard sirens in other frequency bands. Here, we want to emphasize another advantage: Since SMBHBs are all hosted in supermassive galaxies, we can directly exclude those galaxies with low masses when using the dark siren method, and only need to consider the supermassive galaxies within the localization volume of the GW source. Compared with the stellar-mass dark sirens, it can greatly reduce the computational costs and the uncertainties caused by spurious host galaxies in Bayesian inference.




\section{Conclusion}\label{sec4}
Several PTA projects have reported substantial evidence for the existence of a SGWB in the nHz band. In addition to SGWB, the GW signals from individual SMBHBs may also be observed in the future. Such individual GW events can be used as dark sirens to constrain cosmological parameters. In the previous studies on dark sirens, the simulations of SMBHBs were based on the M17 model, which cannot fit the observed SGWB signal well. In this paper, we use the C21 model, which can self-consistently predict the local number density of SMBHBs given the amplitude of the SGWB, to investigate the potential of SMBHB dark sirens in measuring the Hubble constant. According to our results, the following conclusions can be drawn:

(i) In the C21 model, a PTA with 100 MSPs over a 10-year observation period could measure $H_0$ with a precision close to 1\%. If the number of MSPs increases from 100 to 200, a 5-year observation could measure $H_0$ with a precision close to 1\%.

(ii) The C21 model predicts a larger number of SMBHBs than the M17 model. The numbers of SMBHBs derived from the M17 and C21 models are 490 and 213, respectively.  


(iii) The SMBHB distribution in the C21 model is derived from the SGWB signal, which is the overall superposition of the GW signals from individual events. Therefore, the mass distribution of SMBHBs predicted by the C21 model is more statistical than those predicted by the M17 model.

Our conclusions indicate that modeling the SMBHB population has a great impact on the analysis of SMBHB dark sirens. In the future, more accurate measurements of the SGWB signal will not only give us a better understanding of the distribution and evolution history of SMBHBs, but also have important implications for measuring cosmological parameters via the dark siren method.

\acknowledgments
We thank Shang-Jie Jin for helpful discussions.
This work was supported by the National SKA Program of China (Grants Nos. 2022SKA0110200 and 2022SKA0110203), the National Natural Science Foundation of China (Grants Nos. 12473001, 11975072, 11875102, 11835009, 12305069, and 12305058), the National 111 Project (Grant No. B16009), the Program of the Education Department of Liaoning Province (Grant No. JYTMS20231695), and the Natural Science Foundation of Hainan Province (Grant No. 4240N215).

\bibliography{pta}{}
\bibliographystyle{JHEP}

\end{document}